\documentclass[referee,a4paper,12pt,traditabstract]{swsc} 

\usepackage{graphicx}
\usepackage{txfonts}
\usepackage{subfigure}
\usepackage{epstopdf}
\usepackage{lineno}
\usepackage[authoryear,round]{natbib}
\usepackage[backref]{hyperref}
\usepackage{url}

\hypersetup{colorlinks=true,citecolor=cyan,urlcolor=cyan,linkcolor=blue}

\begin{document}


   \title{Climate, weather, space weather: model development in an operational context}

   \titlerunning{Climate, weather, space weather: model development in operational context}

   \authorrunning{Folini}

   \author{Doris Folini
          \inst{1,2}
          }

   \institute{ETH Zurich, Institute for Atmospheric and Climate Science, Zurich, Switzerland\\
              \email{\href{mailto:Doris.Folini@env.ethz.ch}{Doris.Folini@env.ethz.ch}}
         \and
             \'{E}cole Normale Sup\'{e}rieure, Lyon, CRAL, UMR CNRS 5574, Universit\'{e} de Lyon, France
             }


 
             \abstract {Aspects of operational modeling for climate,
               weather, and space weather forecasts are contrasted,
               with a particular focus on the somewhat conflicting
               demands of 'operational stability' versus 'dynamic
               development' of the involved models. Some common key
               elements are identified, indicating potential for
               fruitful exchange across communities.  Operational
               model development is compelling, driven by factors that
               broadly fall into four categories: model skill, basic
               physics, advances in computer architecture, and new
               aspects to be covered, from costumer needs over physics
               to observational data. Evaluation of model skill as
               part of the operational chain goes beyond an automated
               skill score. Permanent interaction between 'pure
               research' and 'operational forecast' people is
               beneficial to both sides. This includes joint model
               development projects, although ultimate responsibility
               for the operational code remains with the forecast
               provider. The pace of model development reflects
               operational lead times. The points are illustrated with
               selected examples, many of which reflect the author's
               background and personal contacts, notably with the
               Swiss Weather Service and the Max Planck Institute for
               Meteorology, Hamburg, Germany. In view of current and
               future challenges, large collaborations covering a
               range of expertise are a must - within and across
               climate, weather, and space weather. To profit from and
               cope with the rapid progress of computer architectures,
               supercompute centers must form part of the team.}


   \maketitle
%
%
\section{Introduction}
\label{sec:intro}
Model development in the context of an operational chain or forecast
service implies that one has to deal with the somewhat conflicting
demands of 'operational stability' versus 'dynamic development'.  On
the one hand, one does not want to touch the operational  model,
whose compliance with specifications has been ascertained - a
state loosely referred to as 'operational stability' in the
following. Specifications typically depend on the forecast service and
cover aspects from false alarm rate over time-criticality of model run
time to compatibility with the entire operational chain. On the other
hand, model development is compelling for a number of reasons,
to be detailed later.  Such 'dynamic development' typically breaks the
aforementioned compliance with specifications, which ultimately has to
be re-established to regain 'operational stability'.  Although a topic
of discussions and meetings, the subject gets rather little coverage
in the literature~\citep[][]{steenburgh-et-al:14}.

The present paper takes a cross-community view on the question, from
the perspective of climate, weather, and space weather. Earlier
studies demonstrated the potential for mutual learning across these
three communities~\citep[][]{sisco:07}. All three of them aim at
predicting the future physical state of the 'Sun-Earth' system in a
likelihood sense and rely on the same, basic building blocks for their
models: physical insight and associated equations - from empirical
dependencies to basic physical laws - that characterize the time
evolution of the system, numerical simulations to integrate the
physical equations in time, and measurement data for model
initialization (often via advanced data assimilation techniques) and
validation~\citep[][]{tsagouri-et-al:13, bauer-et-al:15,
  palmer:16}. To translate the output of the numerical model into
customer specific products, expert knowledge (forecasters) and
dedicated back end models are essential. For example, in the
context of emergency preparedness in case of a nuclear power plant
accident, a Lagrangian particle dispersion model may be used as back
end model to translate the numerical weather prediction (NWP) into a
fall-out prediction~\citep{szintai-et-al:09}. All is time critical,
the prediction must be available before the real event. Differences
among the communities include the envisaged lead time, from minutes to
centuries, the availability of measurement data for model
initialization and validation, or the interest given to extreme
events.

Customer needs - more accurate and comprehensive forecasts - may be
seen as the ultimate driver behind the further development of
operational forecast models. Specific drivers of development are
suggested to fall into four broad categories: improvement of model
skill, insight from basic science, exploitation of new computer
architectures, and coverage of entirely new aspects, e.g. an
additional physical model, new observational data, or customer
needs. Clearly, these four categories are not completely independent
of each other. For example, covering additional physics and
transitioning to a more powerful computer architecture may go hand in
hand. The different drivers reflect in the emerging projects in the
form of project size, duration, or composition of the development
team. Their relative importance depends on the community. Regular
examination of driving factors ideally forms part of the operational
context or even, in the case of model skill, of the forecast
operational chain. Permanent exchange with basic research outside the
operational service is an asset in this process: it provides another
perspective, potentially taps complementary expertise, and enables
co-development projects that translate driving factors into new
(operational) code. The ultimate responsibility for the operational
code (coding rules, documentation, verification, validation etc.)
resides, however, with the operational service institution. It is
argued that such exchange is beneficial to all parties involved. On
the side of research, benefits lie with the use of the operational
code as a clean, well documented and well tested starting point for
research or a showcase application demonstrating the power of new
technologies. The paper aims at illustrating these points for each
community and at carving out the potential for mutual learning across
communities.

Section~\ref{sec:opasp} takes a community specific point of view,
likely with some bias to the author's own background in climate,
weather, and astrophysics, especially when it comes to concrete
examples, which are often from the Swiss Weather Service (MeteoSwiss)
and the Max Planck Institute for Meteorology, Hamburg, Germany. Some
aspects of the operational chain and service are addressed, thereby
embedding the focal point of the paper: how different communities cope
with the challenge of 'operational stability' versus 'dynamic
development'.  Section~\ref{sec:lessons} deals with similarities
across communities and what the different communities might learn from
each other. Conclusions are presented in
Section~\ref{sec:conclusions}.
\section{Operational predictions and development in different communities}
\label{sec:opasp}
The objective of this section is to provide some characterization of
the three communities of interest - NWP in Section~\ref{sec:nwp},
climate in Section~\ref{sec:climate}, and space weather prediction
(SWP) in Section~\ref{sec:spaceweather}. Each of these three sections
is structured around roughly the same basic points. The goals of each
community are sketched, including the targeted temporal and spatial
scales. Aspects of associated operational modeling are given, from
institutional issues to assessment of model skill. Model development
within this context is then illustrated, form factors triggering
development projects, over how such projects work, to integration of
the new development into the operational model.  Some examples are
given.
\subsection{Weather Prediction}
\label{sec:nwp}
\subsubsection{Objectives}
Operational NWP deals with lead times of hours to days and seasons
(with decadal predictions emerging) on global to regional to local
scales~\citep{meehl-et-al:14}. Specific back-end codes that run after
the NWP model are used to meet demands from a wide range of customers,
from governments (e.g. flight safety, hurricanes) over business
(e.g. tourism, water management) to individuals (e.g. pollen
forecast). Interests are with both, 'ordinary weather' and extreme
events.
\subsubsection{Modeling Aspects}
The range of involved scales, from global to local, translates into
running a hierarchy of models: a global scale model (few ten kilometer
grid cell size) provides the large scale dynamics and 'long term
perspective', which are used as boundary conditions for a more finely
resolved (kilometer scale) regional model or a hierarchy
thereof. Coupling across scales is one way (from coarse to fine) or,
more recently, two-way (from coarse to fine and from fine to
coarse)~\citep[e.g.][]{reinert-et-al:17}. It may involve the same
model at different resolutions, or different models (see below).

For the time scales of interest, initial conditions play a crucial
role. Consequently, much effort has gone in model initialization
techniques that furnish initial conditions close to observations /
reality, yet compatible with the physics covered by the (imperfect)
model~\citep[e.g. 4D Var, ][]{courtier-et-al:94, tremolet:06,
  zhang-pu:10, shaw-daescu:17}. To address the issue of weather being
a chaotic system~\citep{lorenz:63}, modern weather forecasts provide
likelihoods. This means that not one simulation but an ensemble of
simulations differing in their initial conditions must be run in a
time-critical fashion - the prediction must be available way ahead of
the predicted event. More recently, ensembles are also being used to
explicitly sample model uncertainty, for example due to the choice of
numerical values in sub-grid-scale
parameterizations~\citep[e.g.][]{leutbecher-et-al:17}. Ensemble
predictions provide a framework to extend the forecast lead time in a
meaningful way~\citep{buizza-leutbecher:15}. But they are also a
significant cpu cost factor~\citep[e.g.][]{leutbecher-et-al:17}.
Similarly, finer grid resolution increases the cpu costs, yet is
highly desirable to better capture relevant small scale features, like
mountains or coast-lines, and to resolve physical processes that
otherwise have to be included via sub-grid scale parameterization,
e.g. convection and associated
precipitation~\citep[e.g.][]{langhans-et-al:12, ban-et-al:14}.

The above demands - hierarchy of models, observation based model
initialization, ensemble prediction, all time-critical - necessitate
NWP and associated model development to be an overall highly
collaborative effort. This although individual weather services
typically run an operational model / chain themselves. An impression
can be obtained from the project web-pages of some regional scale
consortia (COnsortium for Small-scale Modeling, COSMO,
http://www.cosmo-model.org/; High Resolution Limited Area Model,
HIRLAM, http://hirlam.org/; the Weather Research Forecasting model,
WRF, http://www.wrf-model.org). From these consortia it also becomes
obvious that hierarchies often go not only across models but also
across institutions (e.g. global from the European Center for
Medium-Range Weather Forecasts, ECMWF, regional from COSMO), making
clean interfaces a must. The latter enable a comparatively easy hand
shake between different models or also models and observations or
models and costumers. The concrete meaning of the term interface is
correspondingly broad, ranging from definitions for observations
(what, when, how), over unit conventions and coordinate system
definitions to data formats (netcdf and grib) and detailed
documentations of data and model. Some of these standards have been
put forward by the World Meteorological Organization (WMO,
http://www.wmo.int). Others may be seen more as the result of
co-evolution among different stake-holders. With regard to the model
as such, the large consortia and the strict rules (e.g. coding but
also licensing issues) that have to go with them add to the basic
challenge of 'operational stability' versus 'dynamic development'.
\subsubsection{Model Development}
A main development driver in NWP is the permanent evaluation of
forecast skill. This is possible because of the short lead times,
because interest is also with ordinary weather, i.e., each forecast
and not only extreme events, and because there is ample observational
data at hand on Earth. Comparison of forecast and reality is
typically done via an institution specific skill metric. Other
measures may be added on top. For example, as part of the operational
chain at MeteoSwiss forecasters meet after each shift, while memory is
still fresh and as part of the operational chain, with people from
modeling to provide their impression of the model's performance for
the last forecasting period (personal communication P. Steiner,
MeteoSwiss). Several times a year this information is analyzed for
model deficiencies that escaped the automated skill score.

Model improvement can then come via an internal project at the weather
service or via a joint project with universities~\citep[e.g. improved
terrain following coordinates, ][]{schaer-et-al:02}. On the university
side, a motivation for such a joint project is the free use of most or
all of the operational NWP codes, tools, and data for their
research. For the weather service, this form of exchange allows to
carry out also larger exploratory projects with a close link to basic
atmospheric science, which might be difficult to realize otherwise,
e.g. for lack of expert knowledge or funding possibilities. Such
projects may last anywhere from months to years and may cover anything
from small adaptations to adding new capabilities / components to the
model. If they lead to improved predictions within the research
context, they are professionally implemented and tested (verification,
validation, calibration / tuning, effect on customer specific
back-ends) by the weather service, following strict rules and
protocols. If all tests are passed, the development enters the
operational code. Corresponding releases take place several times per
year at MeteoSwiss. Note that this implies that forecasters and
customers must cope with frequent - albeit typically small - changes
of the product they get.

Another development driver may be summarized as insight from
basic research. Into this category falls the awareness that ensemble
simulations can be used to translate imperfections in initial
conditions and model formulation into meaningful probabilistic
forecasts. Associated development projects tend to be rather long
and complex.

Yet another development driver is the advance of computer
architectures. The reward is, roughly speaking, more computation in
shorter time, potentially even for less energy. In terms of the above
examples, it enables finer resolution and larger ensembles. This
potential has spurred interest in porting (operational) NWP to modern
computer architectures, in particular, graphical processing units /
GPUs~\citep{michalakes-vachharajani:08, shimokawabe-et-al:14,
  vanderbauwhede-takemi:16, deconinck-et-al:17}, but there are also
attempts toward using cloud computing~\citep{molthan-et-al:15,
  siuta-et-al:17, blaylock-et-al:17, chen-et-al:17}. Associated
challenges comprise scalability (the model should run on thousands of
cores), fault awareness and tolerance (the model should tolerate
failure of a thread or core), advanced data compression techniques and
I/O, or also concepts for portability (computer architectures may
differ among consortia members) and composability (relevant physics
may differ among users, e.g.  mountain snow pack or storm surges) or
modularity~\citep{palmer:15, bauer-et-al:15,
  dueben-dawson:17}. Tackling these challenges requires - besides
domain scientists - highly specialized knowledge on computing related
topics, ideally in close collaboration with a large supercompute
center, several years time, and likely some shift of paradigms. Among
the latter are, for example, the question of programming language
(replacement of fortran, at least in parts), the requirement of
bit-reproducibility, partial use of reduced precision
arithmetic~\citep{dueben-et-al:14}, online analysis / re-calculation
to reduce I/O, or whether domain scientists are willing to give up
control over parts of the model (e.g. solution of a Laplacian) in
favor of domain specific languages or computer architecture specific
libraries.

On the positive side, the range of stake-holders involved can also
open additional funding opportunities for development projects.
Recent example of this kind are the project on Energy-efficient
Scalable Algorithms for Weather Prediction at Exascale (ESCAPE,
http://www.hpc-escape.eu/) or also the Center of Excellence in
Simulation of Weather and Climate in Europe (ESiWACE,
https://www.esiwace.eu/). Both initiatives explicitly address both,
NWP and climate. In addition, there are many smaller initiatives
e.g. within the context of Partnership for Advanced Computing in
Europe (PRACE, http://www.prace-ri.eu or). So far, to the author's
knowledge, only one operational NWP model has been successfully ported
to GPUs, namely that of MeteoSwiss~\citep[e.g.][]{fuhrer-et-al:14,
  gysi-et-al:15, leutwyler-et-al:16, prein-et-al:17}. The
corresponding development branch of the code was largely decoupled
from the operational version for years. The effort turned out to be a
win-win situation for all partners: a more powerful yet less energy
intense code for operational forecasts and research, as well as a
showcase application that highlights the possibilities offered by new
computer architectures at the Swiss National Supercompute Center
(CSCS).
\subsection{Climate Projections}
\label{sec:climate}
\subsubsection{Objectives}
Climate projections aim at lead times of decades to centuries, at
global to regional spatial scales. The term 'operational' is hardly
used in the context of climate modeling. Yet the model data entering
the International Panel for Climate Change (IPCC)
reports~\citep{ipcc:15}, for example the global climate model data
from the Coupled Model Intercomparison Project Phase 5 (CMIP5),
fulfill some operational key characteristics: the data are to some
degree customer driven (United Nations Environment Program, UNEP;
World Meteorological Organization, WMO; governments or business for
selected back-end products), have to be available in time (for the
IPCC report), and respect a number of specifications. The
specifications arise from the desire to be able to compare and re-use
the model data submitted to IPCC and are defined by working groups
from the climate modeling community as such~\citep[for CMIP,
see][]{taylor-et-al:12,
  eyring-et-al:16}. Broadly speaking, they detail the setup of some
common simulations and of some aspects of the model. An example for
the former is the demand to perform a simulation covering the years
1850 to 2005 with predefined input data, like annual mean greenhouse
gas concentrations. Model specifications include that a certain amount
of physical components must be covered (e.g. atmosphere, ocean) and
that standards enabling data exchange be respected (e.g. file format,
unit conventions, coordinate system definitions, variable names).
From this perspective, climate projections also face the issue of
'operational stability' versus 'dynamic development' and accompanying
effects on the development process~\citep{jakob:15}.

The term 'operational' refers, however, to a quite different product
than in NWP, in terms of lead time and spatio-temporal scales, but
also concerning product details - rather customer tailored in the case
of NWP, closer to the model output as such in the case of
climate. Associated are differences in terms of operational model
(e.g. physical components covered), embedding into an operational
chain (more elaborate in NWP), bodies behind the operational product
(larger in NWP), and model development. The points are further
illustrated in the following, with focus on global climate models
(GCMs) / Earth System Models (ESMs) and CMIP5. Similar
considerations apply to regional climate models.
\subsubsection{Modeling Aspects}
The long lead times are a distinguishing feature of (operational)
climate modeling. They have several consequences for the design,
operation, and further development of such models. First, they require
ESMs to take into account additional system components besides the
planetary atmosphere, for example oceans, sea-ice, or vegetation. The
exchange between corresponding model components (or models, for short)
is mostly two-way, i.e., information (e.g. an energy flux) is passed
from one model to another and vice versa. Developments on one model
thereby tend to impact others. Model components (e.g. atmosphere or
ocean) are typically developed and brought to operational stability on
a stand alone basis, before being coupled and finally adjusted across
components~\citep[see e.g. ][]{hourdin-et-al:17}. Second, the long
lead times allow for and also demand for longer model run times, from
about a week to several months for one simulation. Third, data
assimilation for initialization is less of an issue than in NWP, as
either the system memory is much shorter than the lead time (e.g. for
the atmosphere) or because comparatively little observational data for
assimilation is available (e.g. for the deep ocean). To nevertheless
arrive at a controlled initial state, an ESM is typically relaxed by
running it for several thousand years with fixed setup,
e.g. conditions as of the year 1850. Fourth, to arrive at a controlled
initial state that is compatible with observed climate variables
(e.g. a global mean temperature remaining around 13.7 degree Celsius
in 1850), numerical values in sub-grid scale parameterizations have to
be adjusted: the model has to be calibrated or
tuned~\citep[e.g.][]{schmidt-et-al:17, hourdin-et-al:17}. Because of
the long time scales involved (run time and system memory) this
process takes from several months to over a year and consumes a
significant amount of cpu. The calibration and relaxation process, as
well as the generally long integration times for actual operational
production, reflect in comparatively long (years) intervals between
individual operational code versions. The production phase of one
operational version typically coincides with the development phase of
the subsequent model version and resources (people, cpu) have to be
split between the two.

Operational climate modeling and associated model development is
essentially shouldered by individual research institutions (around
thirty in the case of the last IPCC report / CMIP5, often in a stable
team with some supercompute center). This is in contrast to NWP, where
large, often transnational consortia play an important role (see
Section~\ref{sec:nwp}). The institution based approach has the
advantage that the exchange between development and operation is
typically easier. Operational and research codes are closely related,
the latter just being (research purpose adapted) branches of the
former. In this way, research can rely on the operational code as a
solid basis that continues to be well tested by research and
operational activities. Basic research developments thus are made
already within the context of the operational code, although often not
in 'operational coding quality'. On the downside, an institution
shouldering both, research and operational modeling, must split its
resources, like people or available cpu time, between the two.  At
times, this may result in operational goals (like IPCC) being
prioritized over 'pure research', one reason being that the former
tend to be more time critical than the latter.  Corresponding concerns
were raised, for example, during a recent workshop on Earth System
Modeling (https://www.4icesm.eu).

Rules and regulations associated with the operational model are mostly
institution specific as well, except for specifications regarding data
exchange (see above). They are a must given typical code sizes, and
range from coding rules over version control (e.g. svn, git) to
validation / model skill (see below). Yet they tend to be less strict
and comprehensive than in NWP, possibly reflecting the different sizes
of the communities behind a single model. NWP likely also requires
more regulations as the operational model forms part of a complex
operational chain, from data assimilation to user specific back end
models and products (see Section~\ref{sec:nwp}), while climate
projections are more stand alone. They rely on comparatively simple
input data (no data assimilation) and provide essentially the model
output as such (no user specific back end products or models).
\subsubsection{Model Development}
With climate research and operational climate modeling coming out of
essentially one hand, much model development simply results from
research projects - except for the operational implementation itself,
which is often done by a few, specialized people.

A main development driver is again model skill, now typically with
regard to observation based climatological mean quantities, possibly
their variance, like (global mean) temperature, the Indian monsoon, or
El Ni\~{n}o. Corresponding skill metrics are typically institution
specific. A common skill metric that would allow to compare model
skill across institutions is suggested for
CMIP6~\citep{eyring-et-al:16}. Comparison of observed and modeled
historical climate, roughly from 1850 till 2000, is more of a final
(yet crucial) test for an ESM than a primary driver of its
development~\citep{hourdin-et-al:17}. Translating a lack of model
skill into a concrete, targeted development project tends, however, to
be even more challenging in an ESM (with atmosphere, ocean, sea ice,
vegetation, etc.)  than in NWP.

An emerging field of climate model development is the use of an ESM in
NWP mode and testing its skill
correspondingly~\citep[e.g.][]{simmons-et-al:16}. The rational behind
is that climate may be seen as the long-term statistical average of
weather, thus a good climate model should also have good NWP
properties. As the atmospheric component of the ESM is tested against
more detailed observational data than the above mentioned
climatological mean quantities, it is put on more firm physical
ground. The approach is also a step toward bridging the lead time gap
between NWP and climate, thus toward seamless
predictions~\citep[][]{palmer-et-al:08, simmons-et-al:16}.

A lasting driver of model development in climate is the coverage of
additional system aspects, like for example replacing prescribed
plants and carbon sources / sinks with a carbon cycle model comprising
interactive vegetation (e.g. trees growing or dying), carbon storage
in oceans, etc.  Associated stand alone models may originate from pure
research. Once such a model seems mature enough, in terms of science
and efficiency of execution / cpu requirements, one may try to couple
this model as yet another component to the ESM. Keeping ESMs modular
such that they allow for easy coupling of new stand alone models is an
endeavor. It is also a challenge with regard to coding, including the
potentially conflicting demands of modularity on the one hand and
optimization for execution speed on the other hand.

Basic physics in the sense of switching to more physically sound
sub-grid-scale parameterization is another important driver. However,
within the operational context this driver has the downside that
physically improved parameterization may first result in reduced model
skill, as errors that used to compensate no longer do
so~\citep{jakob:15, hourdin-et-al:17}. Additional development time is
needed, which is potentially (too) costly for an individual
institution. Similar considerations apply with regard to exploitation
of new computer architectures. Promises lie, for example, with better
spatial resolution to address regional projections and the role of
clouds. Challenges include lack of man power, large codes, code
portability, and uncertainty about the longevity of different
accelerator technologies - thus about whether portation would pay off
at all.

The situation may change as climate and weather models approach and
synergies can be exploited in larger teams. An example in this
direction is the ICOsahedral Non-hydrostatic model (ICON) used by the
German Weather Service as operational global NWP model and by the Max
Planck Institute for Meteorology, Hamburg, Germany as global climate
model. Together with other stake-holders from NWP, supercompute centers,
and universities, several joint projects are under way to explore
portation of ICON to new computer architectures (e.g. within the
Platform for Advanced Scientific Computing, PASC,
http://www.pasc-ch.org/).

In summary, developments in an operational ESM context go at a much
slower pace than NWP development just because of the physical and lead
time scales involved in the problem. However, the number of people
involved in model development likely also plays a role. Whether larger
collaborations would make the development process more efficient and
faster is a matter of debate~\citep{palmer:16}.  One danger to ever
growing collaborations is a loss of diversity.
\subsection{Space Weather Prediction}
\label{sec:spaceweather}
\subsubsection{Objectives}
Space weather prediction (SWP) deals with perturbations originating at
the Sun - eruptive perturbations, notably X-ray flares, coronal mass
ejections (CMEs), and solar energetic particle events (SEPs), as well
as more persistent features, notably coronal holes and associated wind
streams - and their propagation toward and effects at or near the
Earth. Customer specific interests include satellite safety, high
frequency communication black outs, or damages to electrical power
grids from geomagnetically induced
currents~\citep[e.g.][]{sibley-et-al:12, riley-et-al:18}. Lead times
depend on the type of event and typically range from minutes to days,
although longer time scales are also of
interest~\citep[e.g. recurrence of coronal hole associated with the
solar rotation period of 27 days; the solar cycle of about 11 and 22
years][]{watermann-et-al2:09, singh-et-al:10}.  Operational products
range from publicly available activity indices to customer tailored
quantities~\citep[see e.g., ][]{araujo-pradere:09, tsagouri-et-al:13,
  steenburgh-et-al:14, schrijver-et-al:15, bonadonna:17}.
\subsubsection{Modeling Aspects}
Operational space weather providers include consortia (e.g. the teams
participating in the ESA Space Situational Awareness (SSA) Programme's
Space Weather Service Network, http://swe.ssa.esa.int/) and
organizations engaged in NWP (e.g. the Space Weather Prediction
Center, SWPC, of the National Oceanic and Atmospheric Administration,
NOAA, http://www.swpc.noaa.gov). New developments in space weather
services are taking place in a number of countries, e.g. the United
Kingdom, Belgium, Poland, Sweden, Austria, Australia, Brazil, Mexico,
Canada, Korea, Japan, China, Indonesia, India, or also South Africa. A
collaborative network of space weather service-providing organizations
around the globe is provided by the International Space Environment
Service (ISES). The ISES mission is to improve, to coordinate, and to
deliver operational space weather services through a network of
Regional Warning Centers (http://www.spaceweather.org/)

The different types of perturbations (X-ray flares, SEPs, CMEs,
coronal holes) find their correspondence in rather separated modeling
communities~\citep[][]{zhao-dryer:14, luhmann-et-al:15,
  reiss-et-al:16, barnes-et-al:16, cranmer-et-al:17,
  murray-et-al:17}. Further splitting of modeling activity occurs for
regions closer to Earth (magnetosphere, ionosphere / thermosphere,
Earth atmosphere and surface) because of traditional scientific
domains, specific customer needs, as well as the physical processes
involved~\citep[][]{lathuillere-et-al:02}. Models range from empirical
to semi-empirical to physics based.  An impression of the emerging,
rather fragmented modeling landscape may be obtained from the SWPC
web-page. A number of projects aim at combining this existing
expertise to arrive at more comprehensive space weather models. This
includes coupling of different models, i.e., using the output of one
model as input / initialization of another model. The coupling is
typically one way, with information being passed from Sun to
Earth. Concrete initiatives include the Space Weather Modeling
Framework (SWMF) at the University of
Michigan~\citep[http://csem.engin.umich.edu/][]{toth-et-al:05,
  toth-et-al:12} and the Virtual Space Weather Modeling Center (VSWMC,
https://esa-vswmc.eu/). A complementary, more integrative approach
that stresses the critical linking of multiple scales at shocks,
interfaces, and reconnection sites, is taken by the Space Weather
Integrated Forecasting Framework~\citep[SWIFF,
http://www.swiff.eu/][]{lapenta-et-al:13}.

Model initialization relies on satellite or other observational data
and uses a range of data assimilation techniques~\citep[e.g. Ensemble
Kalman Filtering or Ensemble Optimal Interpolation
methods][]{hickmann-et-al:15, murray-et-al:15}. Data assimilation for
model initialization is, however, not as widely spread in SWP as in
NWP.  Several reasons for this difference are identified
by~\citet{lang-et-al:17}, a major one being data
availability. Translating the uncertainty from initialization into
probabilistic forecasts using different ensemble
techniques~\citep[][]{schunk-et-al:14, elvidge-et-al:16, knipp:16,
  owens-et-al:17} is getting standard. Dependence on initial
conditions can be chaotic~\citep[as in NWP, e.g. magnetosphere /
ionosphere / thermosphere][]{horton-et-al:01, mannucci-et-al:16,
  wang-et-al-2:16} or non-chaotic~\citep[e.g. CME propagation toward
Earth][]{lee-et-al:13, lee-et-al:15, cash-et-al:15,
  pizzo-et-al:15}. In the later case, the accuracy of a prediction
will strongly depend on the quality of the underlying initial data,
for example the parameters of a CME and the characterization of the
solar wind to be passed. Depending on how model skill is evaluated it
may then rather be an evaluation of 'initial condition skill' than of
actual model skill. Also, one model may outperform another because it
was designed to take the imperfection of the initial data into
account.
\subsubsection{Model Development}
Regarding model development, the relative importance of (observation
based) model initialization as well as the many existing, largely
stand alone models allow for development of individual models without
affecting others. Together with the short lead times, thus short run
times, this potentially allows for overall short development
cycles. An interesting account of a concrete development project in
the context of operational SWP is given
by~\citet{steenburgh-et-al:14}. It addresses a wider scope than the
present paper, touching for example also on visualization tools for
the model results. With the rich and detailed presentation of real
world issues that have to be dealt with to make a model operational,
the paper makes an ideal, complementary reading to the present
paper. Like the present paper, it stresses the importance of close
exchange between research and operational people.

Regarding drivers of model development, model skill is again a
prominent driver, despite the complicating aspects mentioned above.
For the many empirical models used in and specifically designed for
SWP, model skill may be the single most important development
driver. Evaluation of model skill comes in the form of automated,
model specific skill metrics, but also in at least some institutions
in a 'soft variant' via regular meetings between modelers and
forecasters~\citep{steenburgh-et-al:14}.  Different actors in
(operational) SWP tend to use different measures of model skill, there
is no wide acceptance of a best approach. There are, however,
initiatives toward more unified and thus comparable model skill
evaluation. The ISES network as well as the ESA SSA Space Weather
Service Network, both mentioned already earlier, engage in this
direction. Efforts by the Community Coordinated Modeling Center (CCMC,
https://ccmc.gsfc.nasa.gov/) build on simulations of the same, real
events and use of the same skill metric for all comparable models. A
corresponding platform for the comparison of real-time forecasts, the
predictions being uploaded to CCMC by different providers, is
operational for CMEs, under implementation for flares, and planned for
SEPs. The approach is interesting as it allows to identify systematic
deficiencies across models. In practical terms this also means
assistance of community wide model validation
efforts~\citep[e.g.][]{pulkkinen-et-al:13, rastaetter-et-al:14,
  glocer-et-al:16, welling-et-al:17}. SWP here is a trend setter.  In
climate, a prescribed, common skill metric is only suggested for the
next inter-comparison, CMIP6~\citep[][]{eyring-et-al-2:16}. In NWP,
such a common skill metric is less straightforward and maybe less
appropriate, as relevant weather characteristics are rather regional
and very different in, say, Central Europe and Southern India. Common
are, however, the concepts and ideas behind the skill metric. All
three communities thus may profit from associated guidelines and
recommendations put forward by the WMO Joint Working Group on Forecast
Verification Research (JWGFVR).

Basic physics appears a key driver of operational model development
for those models that are physics based and used in both research and
operational SWP~\citep[e.g. WSA-Enlil, see][for an
overview]{mays-et-al:15}. Much like in climate and NWP, the use of the
same model potentially exploits synergies (man power, expertise,
tools, etc.) and inspires improvements also of the operational code
version. The use of a common code facilitates transfer from research
developments into operation. As such a code is potentially used in
widely different environments, e.g. in terms of computer architecture,
operating system or file system, possibly in serial and parallel mode,
this promotes overall robustness and portability of the
code~\citep{steenburgh-et-al:14}

Advances in the form of more, better, and different observational data
are a definite driver of model development in operational SWP. This
applies especially but not exclusively for empirical models, which
rely most heavily on observations for model design and
application. Changes in computer architecture, by contrast, seem less
of a driver~\citep[see e.g.][]{feng-et-al:13}. Potential reasons could
be lack of free resources or that advances are expected rather
from progress in physical understanding and improved models (empirical
and physics based) than from more cpu.

The relative importance of (observation based) model initialization
allows for development of individual models without affecting
others. Together with the short lead times, thus short run times, this
potentially allows for overall short development cycles.
\section{Operational modeling and development - synthesis across communities}
\label{sec:lessons}
The previous section outlined some characteristics of modeling and
model development in an operational context in individual communities,
notably NWP, climate, and SWP. What can be synthesized?

A first impression is that exchange between 'basic research' and
'operational forecasting' ideally is a two-way road with added value
for both sides. From research to the operational side, there is the
asset of enabling exploitation of new scientific developments in an
operational context. Also, there is the research view on what a model
may or may not be able to realistically capture. The opposite
direction - what research might benefit from operational modeling -
is less frequently highlighted. Yet it is common practice, at least
in NWP and climate, that operational codes and even entire modeling
chains or parts thereof are used for basic research. This allows the
latter to build on a code that runs reliably, is well written and
tested (verified and validated), well optimized, and comes with a
version history and a wealth of useful tools, ranging from code
development over validation and visualization to well defined data
formats and interfaces. Also, over time the permanent operational use
exposes potential weaknesses of the model much better than few
dedicated tests in a pure science project. Turning again to the
direction from research to operation, use of the operational code in
research potentially facilitates transfer of improved functionality
back to the operational code. 

A second aspect concerns model skill. Skill metrics and associated
scores are an ubiquitous concept. They are a driver of model
development, although typically a lack of skill not easily translates
into a concrete cause thereof, a concrete development project.  They
are attractive because they are quantitative and reproducible, thus
comparable, and transparent. However, they measure only what they are
designed for, and the concrete design of the metric - what quantity is
evaluated, against which benchmark, and what norm is used, e.g. root
mean square or other - often varies with model and / or institution.
This also because models are always imperfect, and a decision has to
be taken as to which aspect of the model should weigh heavier in the
skill score. To compensate for the first flaw (rigid design), a 'soft
model skill assessment' (exchange between modelers and forecasters)
forms part of the operational chain in some NWP and SWP institutions.
Climate modeling is less subject to this flaw in the first place, as
research and operation are much closer, on a personal level, including
exchange about model performance beyond predefined skill metrics.  The
second flaw (model / institution specific score) is typically
addressed via dedicated model intercomparison projects, e.g. CMIP
mentioned in Section~\ref{sec:climate} or CCMC (see
Section~\ref{sec:spaceweather}). Validation of models against the same
skill metrics and observational data is most valuable as it allows to
identify systematic deficiencies across models.  For regional NWP
model inter-comparison is less practical as models are calibrated for
good performance within their operational region.

Third, there is the interplay between drivers, lead time, and time
scales of model development.  Put simply, short lead times reflect in
short model run times and frequent comparison of 'prediction' and
'reality'. The latter can be an essential driver of model development
in an operational context, provided that enough data for model
evaluation are available. For NWP this is indeed the case, resulting
in frequent (possibly) small changes to the operational model version
(up to several times a year in NWP, see Section~\ref{sec:nwp}).
Whether an equally high update frequency of the operational model is
desirable in SWP is less clear. The short run times would allow for
fast development cycles. The comparatively large number of stand alone
models involved seems of little consequence as they are largely
independent of each other (see Section~\ref{sec:spaceweather}). An
obstacle may be observational data coverage to evaluate (long term)
model performance, especially when it comes to (rare) extreme
events. Also, frequent updates of the operational model imply that
customers have to be prepared for slight but frequent changes of their
products, similar as in the case of NWP (see
Section~\ref{sec:nwp}). Climate modeling has much longer development
cycles. There is no short time development driver in the form of daily
comparison of forecast and reality. And changes in one of the
sub-model components (ocean, atmosphere) can necessitate adjustments
in another component, if only via 're-tuning' the model (see
Section~\ref{sec:climate}).  A consequence of the longer development
cycle is that customer products (e.g. for IPCC) tend to differ
substantially between subsequent operational versions.

Finally, there are common challenges ahead, a major one being advances
in computer architectures. On the positive side, these are an
important factor enabling better models, better predictions. However,
to take advantage of this progress, one has to cope with the fact that
computer architectures, chips, and disks, follow their own evolution.
It seems highly unlikely that research or operational services can
exert any substantial influence on chip or disk manufacturers. Slight
influence on a concrete machine procurement might be taken via the
benchmark applications normally run by a supercompute center before
buying a new machine. How much adaptation then is compelling or rather
a matter of choice - Is a GPU code a must? Is a commercial cloud an
alternative? - is a subject of much debate. Opinions also diverge on
adaptation strategies, one idea being to separate codes into machine
specific back ends and science code front ends. The coding and
maintenance of such back ends beyond show case applications is,
however, an open questions. Whether sufficiently many actors will
embark on this approach and could agree to common back ends, thereby
distributing associated costs, remains to be seen.

What seems certain is that complexity will grow and with it the need
for expert knowledge to develop, optimize, and run corresponding well
structured and modular models. In concert, there will be a growth in
complexity of the hand shake between 'research' and 'operation', as
well as between established categories, such as physics or domain
science, scientific computing, and computational science. A partial
answer to this increasing complexity must lie in education and
community culture, promoting heterogeneous teams that cover all
necessary expertise and interact on an equal footing. An asset for
such teams would be the possibility for science career tracks that are
situated between or even alternate among established categories, have
a long term perspective, and share the recognition of 'pure science'
tracks. Today, such cross-cutting careers are rather exceptional.

With the necessarily growing investments in codes and operational
infrastructure, licensing issues potentially raise in
relevance. Transnational collaborations and the associated mix of
legal systems potentially add to the issue. Whether open source
licensing could be an answer is a matter of debate.
\section{Conclusions}
\label{sec:conclusions}
This paper examined the somewhat conflicting demands of 'operational
stability' versus 'dynamic development' in three different
communities: numerical weather prediction, space weather prediction,
and climate projections. Similarities and differences in how the three
communities deal with this issue were identified. In all three
communities, the lead is with the operational side, which is obliged
or even legally bound to meet customer demands. The research community
plays, however, an important part in an overall win-win situation. To
what degree the latter is indeed recognized and appreciated by all
partners is difficult to judge. Both sides have common goals - learn
about dependencies in the system with the goal of making predictions
and, ultimately, understand the underlying mechanism - and in pursuing
them may benefit from each others strengths: clean, verified, and
validated code with plenty of tools and data on the side of
operational services and additional expertise, man power, time and
funding for exploitative scientific studies on the research
side. Mutual awareness of and respect for each other's strengths and
limitations is essential for success. Leaving the community specific
perspective while writing this paper, my conviction grew that mutual
exchange among NWP, SWP, and climate may not always be easy but has
much potential given the similarities in physics, equations, goals,
and challanges ahead. Some of these challenges - notably the need for
composability in view of modern computer architectures and, more
generally, the growing complexity in operation and research - are not
only common to all three communities but call for common
action. Community building, education, but also career paths must
account for these developments in order to ascertain long-term
success.
\begin{acknowledgements}
  I would like to thank Gianni Lapenta, KU Leuven, for inviting me to
  give a talk at the European Space Weather Week in 2016 on the topic
  of the present paper. Two anonymous referees I would like to thank
  for their most constructive comments. To Christoph Sch\"{a}r and
  Martin Wild, Institute of Atmospheric and Climate Science, ETH
  Zurich, I am grateful for their permanent support, in this and other
  activities. Special thanks go to Philippe Steiner and Pirmin
  Kaufmann from MeteoSwiss for all the extensive discussions about
  NWP. Finally, I would like to thank Micka\"{e}l Melzani, Centre de
  Recherche Astrophysique de Lyon (CRAL), and Mark Dieckmann,
  Link\"{o}ping University, for sharing with me their insight into
  plasma kinetics of magnetic reconnection and shocks. The editor
  thanks two anonymous referees for their assistance in evaluating
  this paper.
\end{acknowledgements}


\bibliography{/home/folini/text/articles/MyBigBib.bib}

\begin{thebibliography}{79}
\providecommand{\natexlab}[1]{#1}
\providecommand{\url}[1]{\texttt{#1}}
\providecommand{\urlprefix}{URL }
\providecommand{\eprint}[2][]{\url{#2}}

\bibitem[{{Araujo-Pradere}(2009)}]{araujo-pradere:09}
{Araujo-Pradere}, E.~A.
\newblock {Transitioning Space Weather Models Into Operations: The Basic
  Building Blocks}.
\newblock \emph{Space Weather}, \textbf{7}, S10006, 2009.
\newblock 10.1029/2009SW000524.

\bibitem[{{Ban} et~al.(2014){Ban}, {Schmidli}, and {Sch{\"a}r}}]{ban-et-al:14}
{Ban}, N., J.~{Schmidli}, and C.~{Sch{\"a}r}.
\newblock {Evaluation of the convection-resolving regional climate modeling
  approach in decade-long simulations}.
\newblock \emph{Journal of Geophysical Research (Atmospheres)}, \textbf{119},
  7889--7907, 2014.
\newblock 10.1002/2014JD021478.

\bibitem[{{Barnes} et~al.(2016){Barnes}, {Leka}, {Schrijver}, {Colak},
  {Qahwaji} et~al.}]{barnes-et-al:16}
{Barnes}, G., K.~D. {Leka}, C.~J. {Schrijver}, T.~{Colak}, R.~{Qahwaji}, et~al.
\newblock {A Comparison of Flare Forecasting Methods. I. Results from the
  All-Clear Workshop}.
\newblock \emph{The Astrophysical Journal}, \textbf{829}, 89, 2016.
\newblock 10.3847/0004-637X/829/2/89, \eprint{1608.06319}.

\bibitem[{{Bauer} et~al.(2015){Bauer}, {Thorpe}, and {Brunet}}]{bauer-et-al:15}
{Bauer}, P., A.~{Thorpe}, and G.~{Brunet}.
\newblock {The quiet revolution of numerical weather prediction}.
\newblock \emph{Nature}, \textbf{525}, 47--55, 2015.
\newblock 10.1038/nature14956.

\bibitem[{{Blaylock} et~al.(2017){Blaylock}, {Horel}, and
  {Liston}}]{blaylock-et-al:17}
{Blaylock}, B.~K., J.~D. {Horel}, and S.~T. {Liston}.
\newblock {Cloud archiving and data mining of High-Resolution Rapid Refresh
  forecast model output}.
\newblock \emph{Computers and Geosciences}, \textbf{109}, 43--50, 2017.
\newblock 10.1016/j.cageo.2017.08.005.

\bibitem[{{Bonadonna} et~al.(2017){Bonadonna}, {Lanzerotti}, and
  {Stailey}}]{bonadonna:17}
{Bonadonna}, M., L.~{Lanzerotti}, and J.~{Stailey}.
\newblock {The National Space Weather Program: Two decades of interagency
  partnership and accomplishments}.
\newblock \emph{Space Weather}, \textbf{15}, 14--25, 2017.
\newblock 10.1002/2016SW001523.

\bibitem[{{Buizza} and {Leutbecher}(2015)}]{buizza-leutbecher:15}
{Buizza}, R., and M.~{Leutbecher}.
\newblock {The forecast skill horizon}.
\newblock \emph{Quarterly Journal of the Royal Meteorological Society},
  \textbf{141}, 3366--3382, 2015.
\newblock 10.1002/qj.2619.

\bibitem[{{Cash} et~al.(2015){Cash}, {Biesecker}, {Pizzo}, {de Koning},
  {Millward}, {Arge}, {Henney}, and {Odstrcil}}]{cash-et-al:15}
{Cash}, M.~D., D.~A. {Biesecker}, V.~{Pizzo}, C.~A. {de Koning}, G.~{Millward},
  C.~N. {Arge}, C.~J. {Henney}, and D.~{Odstrcil}.
\newblock {Ensemble Modeling of the 23 July 2012 Coronal Mass Ejection}.
\newblock \emph{Space Weather}, \textbf{13}, 611--625, 2015.
\newblock 10.1002/2015SW001232.

\bibitem[{Chen et~al.(2017)Chen, Huang, Jiao, Flanner, Raeker, and
  Palen}]{chen-et-al:17}
Chen, X., X.~Huang, C.~Jiao, M.~G. Flanner, T.~Raeker, and B.~Palen.
\newblock Running Climate Model on a Commercial Cloud Computing Environment.
\newblock \emph{Computers and Geosciences}, \textbf{98}(C), 21--25, 2017.
\newblock 10.1016/j.cageo.2016.09.014.

\bibitem[{{Courtier} et~al.(1994){Courtier}, {Th{\'e}paut}, and
  {Hollingsworth}}]{courtier-et-al:94}
{Courtier}, P., J.-N. {Th{\'e}paut}, and A.~{Hollingsworth}.
\newblock {A strategy for operational implementation of 4D-Var, using an
  incremental approach}.
\newblock \emph{Quarterly Journal of the Royal Meteorological Society},
  \textbf{120}, 1367--1387, 1994.
\newblock 10.1002/qj.49712051912.

\bibitem[{{Cranmer} et~al.(2017){Cranmer}, {Gibson}, and
  {Riley}}]{cranmer-et-al:17}
{Cranmer}, S.~R., S.~E. {Gibson}, and P.~{Riley}.
\newblock {Origins of the Ambient Solar Wind: Implications for Space Weather}.
\newblock \emph{Space Science Review}, \textbf{212}, 1345--1384, 2017.
\newblock 10.1007/s11214-017-0416-y.

\bibitem[{{Deconinck} et~al.(2017){Deconinck}, {Bauer}, {Diamantakis},
  {Hamrud}, {K{\"u}hnlein} et~al.}]{deconinck-et-al:17}
{Deconinck}, W., P.~{Bauer}, M.~{Diamantakis}, M.~{Hamrud}, C.~{K{\"u}hnlein},
  et~al.
\newblock {Atlas : A library for numerical weather prediction and climate
  modelling}.
\newblock \emph{Computer Physics Communications}, \textbf{220}, 188--204, 2017.
\newblock 10.1016/j.cpc.2017.07.006.

\bibitem[{{D{\"u}ben} and {Dawson}(2017)}]{dueben-dawson:17}
{D{\"u}ben}, P.~D., and A.~{Dawson}.
\newblock {An approach to secure weather and climate models against hardware
  faults}.
\newblock \emph{Journal of Advances in Modeling Earth Systems}, \textbf{9},
  501--513, 2017.
\newblock 10.1002/2016MS000816.

\bibitem[{{D{\"u}ben} et~al.(2014){D{\"u}ben}, {McNamara}, and
  {Palmer}}]{dueben-et-al:14}
{D{\"u}ben}, P.~D., H.~{McNamara}, and T.~N. {Palmer}.
\newblock {The use of imprecise processing to improve accuracy in weather and
  climate prediction}.
\newblock \emph{Journal of Computational Physics}, \textbf{271}, 2--18, 2014.
\newblock 10.1016/j.jcp.2013.10.042.

\bibitem[{{Elvidge} et~al.(2016){Elvidge}, {Godinez}, and
  {Angling}}]{elvidge-et-al:16}
{Elvidge}, S., H.~C. {Godinez}, and M.~J. {Angling}.
\newblock {Improved forecasting of thermospheric densities using multi-model
  ensembles}.
\newblock \emph{Geoscientific Model Development}, \textbf{9}, 2279--2292, 2016.
\newblock 10.5194/gmd-9-2279-2016.

\bibitem[{{Eyring} et~al.(2016{\natexlab{a}}){Eyring}, {Bony}, {Meehl},
  {Senior}, {Stevens}, {Stouffer}, and {Taylor}}]{eyring-et-al:16}
{Eyring}, V., S.~{Bony}, G.~A. {Meehl}, C.~A. {Senior}, B.~{Stevens}, R.~J.
  {Stouffer}, and K.~E. {Taylor}.
\newblock {Overview of the Coupled Model Intercomparison Project Phase 6
  (CMIP6) experimental design and organization}.
\newblock \emph{Geoscientific Model Development}, \textbf{9}, 1937--1958,
  2016{\natexlab{a}}.
\newblock 10.5194/gmd-9-1937-2016.

\bibitem[{{Eyring} et~al.(2016{\natexlab{b}}){Eyring}, {Righi}, {Lauer},
  {Evaldsson}, {Wenzel} et~al.}]{eyring-et-al-2:16}
{Eyring}, V., M.~{Righi}, A.~{Lauer}, M.~{Evaldsson}, S.~{Wenzel}, et~al.
\newblock {ESMValTool (v1.0) - a community diagnostic and performance metrics
  tool for routine evaluation of Earth system models in CMIP}.
\newblock \emph{Geoscientific Model Development}, \textbf{9}, 1747--1802,
  2016{\natexlab{b}}.
\newblock 10.5194/gmd-9-1747-2016.

\bibitem[{{Feng} et~al.(2013){Feng}, {Zhong}, {Xiang}, and
  {Zhang}}]{feng-et-al:13}
{Feng}, X., D.~{Zhong}, C.~{Xiang}, and Y.~{Zhang}.
\newblock {GPU Computing in Space Weather Modeling}.
\newblock In N.~V. {Pogorelov}, E.~{Audit}, and G.~P. {Zank}, eds., Numerical
  Modeling of Space Plasma Flows (ASTRONUM2012), vol. 474 of \emph{Astronomical
  Society of the Pacific Conference Series}, 131, 2013.

\bibitem[{{Fuhrer} et~al.(2014){Fuhrer}, {Osuna}, {Lapillonne}, {Gysi},
  {Cumming}, {Bianco}, {Arteaga}, and {Schulthess}}]{fuhrer-et-al:14}
{Fuhrer}, O., C.~{Osuna}, X.~{Lapillonne}, T.~{Gysi}, B.~{Cumming},
  M.~{Bianco}, A.~{Arteaga}, and T.~C. {Schulthess}.
\newblock {Towards a performance portable, architecture agnostic implementation
  strategy for weather and climate models}.
\newblock \emph{Supercomputing Frontiers and Innovations}, \textbf{1}, 45--62,
  2014.
\newblock 10.14529/jsfi140103.

\bibitem[{{Glocer} et~al.(2016){Glocer}, {Rast{\"a}tter}, {Kuznetsova},
  {Pulkkinen}, {Singer} et~al.}]{glocer-et-al:16}
{Glocer}, A., L.~{Rast{\"a}tter}, M.~{Kuznetsova}, A.~{Pulkkinen}, H.~J.
  {Singer}, et~al.
\newblock {Community-wide validation of geospace model local K-index
  predictions to support model transition to operations}.
\newblock \emph{Space Weather}, \textbf{14}, 469--480, 2016.
\newblock 10.1002/2016SW001387.

\bibitem[{Gysi et~al.(2015)Gysi, Osuna, Fuhrer, Bianco, and
  Schulthess}]{gysi-et-al:15}
Gysi, T., C.~Osuna, O.~Fuhrer, M.~Bianco, and T.~C. Schulthess.
\newblock STELLA: A Domain-specific Tool for Structured Grid Methods in Weather
  and Climate Models.
\newblock In Proceedings of the International Conference for High Performance
  Computing, Networking, Storage and Analysis, SC '15, 41:1--41:12. ACM, New
  York, NY, USA, 2015.
\newblock ISBN 978-1-4503-3723-6.
\newblock 10.1145/2807591.2807627.

\bibitem[{{Hickmann} et~al.(2015){Hickmann}, {Godinez}, {Henney}, and
  {Arge}}]{hickmann-et-al:15}
{Hickmann}, K.~S., H.~C. {Godinez}, C.~J. {Henney}, and C.~N. {Arge}.
\newblock {Data Assimilation in the ADAPT Photospheric Flux Transport Model}.
\newblock \emph{Solar Physics}, \textbf{290}, 1105--1118, 2015.
\newblock 10.1007/s11207-015-0666-3.

\bibitem[{{Horton} et~al.(2001){Horton}, {Weigel}, and
  {Sprott}}]{horton-et-al:01}
{Horton}, W., R.~S. {Weigel}, and J.~C. {Sprott}.
\newblock {Chaos and the limits of predictability for the solar-wind-driven
  magnetosphere-ionosphere system}.
\newblock \emph{Physics of Plasmas}, \textbf{8}, 2946--2952, 2001.
\newblock 10.1063/1.1371522.

\bibitem[{{Hourdin} et~al.(2017){Hourdin}, {Mauritsen}, {Gettelman}, {Golaz},
  {Balaji} et~al.}]{hourdin-et-al:17}
{Hourdin}, F., T.~{Mauritsen}, A.~{Gettelman}, J.~C. {Golaz}, V.~{Balaji},
  et~al.
\newblock {The Art and Science of Climate Model Tuning}.
\newblock \emph{Bull. Amer. Meteor. Soc.}, \textbf{98}, 589, 2017.
\newblock 10.1175/BAMS-D-15-00135.1.

\bibitem[{IPCC(2013)}]{ipcc:15}
IPCC.
\newblock Climate Change 2013: The Physical Science Basis. Contribution of
  Working Group I to the Fifth Assessment Report of the Intergovernmental Panel
  on Climate Change.
\newblock Cambridge University Press, Cambridge, United Kingdom and New York,
  NY, USA, 2013.
\newblock ISBN ISBN 978-1-107-66182-0.
\newblock 10.1017/CBO9781107415324, \urlprefix\url{www.climatechange2013.org}.

\bibitem[{{Jakob}(2014)}]{jakob:15}
{Jakob}, C.
\newblock {Going back to basics}.
\newblock \emph{Nature Climate Change}, \textbf{4}, 1042--1045, 2014.
\newblock 10.1038/nclimate2445.

\bibitem[{{Knipp}(2016)}]{knipp:16}
{Knipp}, D.~J.
\newblock {Advances in Space Weather Ensemble Forecasting}.
\newblock \emph{Space Weather}, \textbf{14}, 52--53, 2016.
\newblock 10.1002/2016SW001366.

\bibitem[{{Lang} et~al.(2017){Lang}, {Browne}, {van Leeuwen}, and
  {Owens}}]{lang-et-al:17}
{Lang}, M., P.~{Browne}, P.~J. {van Leeuwen}, and M.~{Owens}.
\newblock Data Assimilation in the Solar Wind: Challenges and First Results.
\newblock \emph{Space Weather}, \textbf{15}(11), 1490--1510, 2017.
\newblock 10.1002/2017SW001681.

\bibitem[{{Langhans} et~al.(2012){Langhans}, {Schmidli}, and
  {Sch{\"a}r}}]{langhans-et-al:12}
{Langhans}, W., J.~{Schmidli}, and C.~{Sch{\"a}r}.
\newblock {Bulk Convergence of Cloud-Resolving Simulations of Moist Convection
  over Complex Terrain}.
\newblock \emph{Journal of Atmospheric Sciences}, \textbf{69}, 2207--2228,
  2012.
\newblock 10.1175/JAS-D-11-0252.1.

\bibitem[{{Lapenta} et~al.(2013){Lapenta}, {Pierrard}, {Keppens}, {Markidis},
  {Poedts} et~al.}]{lapenta-et-al:13}
{Lapenta}, G., V.~{Pierrard}, R.~{Keppens}, S.~{Markidis}, S.~{Poedts}, et~al.
\newblock {SWIFF: Space weather integrated forecasting framework}.
\newblock \emph{Journal of Space Weather and Space Climate}, \textbf{3}(27),
  A05, 2013.
\newblock 10.1051/swsc/2013027.

\bibitem[{{Lathuill{\`e}re} et~al.(2002){Lathuill{\`e}re}, {Menvielle},
  {Lilensten}, {Amari}, and {Radicella}}]{lathuillere-et-al:02}
{Lathuill{\`e}re}, C., M.~{Menvielle}, J.~{Lilensten}, T.~{Amari}, and S.~M.
  {Radicella}.
\newblock {From the Sun's atmosphere to the Earth's atmosphere: an overview of
  scientific models available for space weather developments}.
\newblock \emph{Annales Geophysicae}, \textbf{20}, 1081--1104, 2002.
\newblock 10.5194/angeo-20-1081-2002.

\bibitem[{{Lee} et~al.(2015){Lee}, {Arge}, {Odstrcil}, {Millward}, {Pizzo}, and
  {Lugaz}}]{lee-et-al:15}
{Lee}, C.~O., C.~N. {Arge}, D.~{Odstrcil}, G.~{Millward}, V.~{Pizzo}, and
  N.~{Lugaz}.
\newblock {Ensemble Modeling of Successive Halo CMEs: A Case Study}.
\newblock \emph{Solar Physics}, \textbf{290}, 1207--1229, 2015.
\newblock 10.1007/s11207-015-0667-2.

\bibitem[{{Lee} et~al.(2013){Lee}, {Arge}, {Odstr{\v c}il}, {Millward},
  {Pizzo}, {Quinn}, and {Henney}}]{lee-et-al:13}
{Lee}, C.~O., C.~N. {Arge}, D.~{Odstr{\v c}il}, G.~{Millward}, V.~{Pizzo},
  J.~M. {Quinn}, and C.~J. {Henney}.
\newblock {Ensemble Modeling of CME Propagation}.
\newblock \emph{Solar Physics}, \textbf{285}, 349--368, 2013.
\newblock 10.1007/s11207-012-9980-1.

\bibitem[{{Leutbecher} et~al.(2017){Leutbecher}, {Lock}, {Ollinaho}, {Lang},
  {Balsamo} et~al.}]{leutbecher-et-al:17}
{Leutbecher}, M., S.-J. {Lock}, P.~{Ollinaho}, S.~T.~K. {Lang}, G.~{Balsamo},
  et~al.
\newblock {Stochastic representations of model uncertainties at ECMWF: State of
  the art and future vision}.
\newblock \emph{Quarterly Journal of the Royal Meteorological Society}, 2017.
\newblock 10.1002/qj.3094.

\bibitem[{{Leutwyler} et~al.(2016){Leutwyler}, {Fuhrer}, {Lapillonne},
  {L{\"u}thi}, and {Sch{\"a}r}}]{leutwyler-et-al:16}
{Leutwyler}, D., O.~{Fuhrer}, X.~{Lapillonne}, D.~{L{\"u}thi}, and
  C.~{Sch{\"a}r}.
\newblock {Towards European-scale convection-resolving climate simulations with
  GPUs: a study with COSMO 4.19}.
\newblock \emph{Geoscientific Model Development}, \textbf{9}, 3393--3412, 2016.
\newblock 10.5194/gmd-9-3393-2016.

\bibitem[{{Lorenz}(1963)}]{lorenz:63}
{Lorenz}, E.~N.
\newblock {Deterministic Nonperiodic Flow.}
\newblock \emph{Journal of Atmospheric Sciences}, \textbf{20}, 130--148, 1963.
\newblock 10.1175/1520-0469(1963)020<0130:DNF>2.0.CO;2.

\bibitem[{{Luhmann} et~al.(2015){Luhmann}, {Mays}, {Odstrcil}, {Bain}, {Li},
  {Leske}, and {Cohen}}]{luhmann-et-al:15}
{Luhmann}, J., M.~L. {Mays}, D.~{Odstrcil}, H.~{Bain}, Y.~{Li}, R.~{Leske}, and
  C.~{Cohen}.
\newblock {Challenges in Forecasting SEP Events}.
\newblock In AAS/AGU Triennial Earth-Sun Summit, vol.~1 of \emph{AAS/AGU
  Triennial Earth-Sun Summit}, 112.01, 2015.

\bibitem[{{Mannucci} et~al.(2016){Mannucci}, {Hagan}, {Vourlidas}, {Huang},
  {Verkhoglyadova}, and {Deng}}]{mannucci-et-al:16}
{Mannucci}, A.~J., M.~E. {Hagan}, A.~{Vourlidas}, C.~Y. {Huang}, O.~P.
  {Verkhoglyadova}, and Y.~{Deng}.
\newblock {Scientific challenges in thermosphere-ionosphere forecasting -
  conclusions from the October 2014 NASA JPL community workshop}.
\newblock \emph{Journal of Space Weather and Space Climate}, \textbf{6}, E01,
  2016.
\newblock 10.1051/swsc/2016030.

\bibitem[{{Mays} et~al.(2015){Mays}, {Taktakishvili}, {Pulkkinen}, {MacNeice},
  {Rast{\"a}tter} et~al.}]{mays-et-al:15}
{Mays}, M.~L., A.~{Taktakishvili}, A.~{Pulkkinen}, P.~J. {MacNeice},
  L.~{Rast{\"a}tter}, et~al.
\newblock {Ensemble Modeling of CMEs Using the WSA-ENLIL+Cone Model}.
\newblock \emph{Solar Physics}, \textbf{290}, 1775--1814, 2015.
\newblock 10.1007/s11207-015-0692-1.

\bibitem[{{Meehl} et~al.(2014){Meehl}, {Goddard}, {Boer}, {Burgman},
  {Branstator} et~al.}]{meehl-et-al:14}
{Meehl}, G.~A., L.~{Goddard}, G.~{Boer}, R.~{Burgman}, G.~{Branstator}, et~al.
\newblock {Decadal Climate Prediction: An Update from the Trenches}.
\newblock \emph{Bulletin of the American Meteorological Society}, \textbf{95},
  243--267, 2014.
\newblock 10.1175/BAMS-D-12-00241.1.

\bibitem[{{Michalakes} and {Vachharajani}(2008)}]{michalakes-vachharajani:08}
{Michalakes}, J., and M.~{Vachharajani}.
\newblock GPU Acceleration of numerical weather prediction.
\newblock \emph{Parallel Processing Letters}, \textbf{18}(04), 531--548, 2008.
\newblock 10.1142/S0129626408003557.

\bibitem[{{Molthan} et~al.(2015){Molthan}, {Case}, {Venner}, {Schroeder},
  {Checchi}, {Zavodsky}, {Limaye}, and {O'Brien}}]{molthan-et-al:15}
{Molthan}, A.~L., J.~L. {Case}, J.~{Venner}, R.~{Schroeder}, M.~R. {Checchi},
  B.~T. {Zavodsky}, A.~{Limaye}, and R.~G. {O'Brien}.
\newblock {Clouds in the Cloud: Weather Forecasts and Applications within Cloud
  Computing Environments}.
\newblock \emph{Bulletin of the American Meteorological Society}, \textbf{96},
  1369--1379, 2015.
\newblock 10.1175/BAMS-D-14-00013.1.

\bibitem[{{Murray} et~al.(2017){Murray}, {Bingham}, {Sharpe}, and
  {Jackson}}]{murray-et-al:17}
{Murray}, S.~A., S.~{Bingham}, M.~{Sharpe}, and D.~R. {Jackson}.
\newblock {Flare forecasting at the Met Office Space Weather Operations
  Centre}.
\newblock \emph{Space Weather}, \textbf{15}, 577--588, 2017.
\newblock 10.1002/2016SW001579, \eprint{1703.06754}.

\bibitem[{{Murray} et~al.(2015){Murray}, {Henley}, {Jackson}, and
  {Bruinsma}}]{murray-et-al:15}
{Murray}, S.~A., E.~M. {Henley}, D.~R. {Jackson}, and S.~L. {Bruinsma}.
\newblock {Assessing the performance of thermospheric modeling with data
  assimilation throughout solar cycles 23 and 24}.
\newblock \emph{Space Weather}, \textbf{13}, 220--232, 2015.
\newblock 10.1002/2015SW001163.

\bibitem[{{Owens} et~al.(2017){Owens}, {Riley}, and {Horbury}}]{owens-et-al:17}
{Owens}, M.~J., P.~{Riley}, and T.~S. {Horbury}.
\newblock {Probabilistic Solar Wind and Geomagnetic Forecasting Using an
  Analogue Ensemble or ``Similar Day'' Approach}.
\newblock \emph{Solar Physics}, \textbf{292}, 69, 2017.
\newblock 10.1007/s11207-017-1090-7.

\bibitem[{{Palmer}(2015)}]{palmer:15}
{Palmer}, T.
\newblock {Modelling: Build imprecise supercomputers}.
\newblock \emph{Nature}, \textbf{526}, 32--33, 2015.
\newblock 10.1038/526032a.

\bibitem[{{Palmer}(2016)}]{palmer:16}
{Palmer}, T.~N.
\newblock {A personal perspective on modelling the climate system}.
\newblock \emph{Proceedings of the Royal Society of London Series A},
  \textbf{472}, 20150,772, 2016.
\newblock 10.1098/rspa.2015.0772.

\bibitem[{{Palmer} et~al.(2008){Palmer}, {Doblas-Reyes}, {Weisheimer}, and
  {Rodwell}}]{palmer-et-al:08}
{Palmer}, T.~N., F.~J. {Doblas-Reyes}, A.~{Weisheimer}, and M.~J. {Rodwell}.
\newblock {Toward Seamless Prediction: Calibration of Climate Change
  Projections Using Seasonal Forecasts}.
\newblock \emph{Bulletin of the American Meteorological Society}, \textbf{89},
  459, 2008.
\newblock 10.1175/BAMS-89-4-459.

\bibitem[{{Pizzo} et~al.(2015){Pizzo}, {de Koning}, {Cash}, {Millward},
  {Biesecker}, {Puga}, {Codrescu}, and {Odstrcil}}]{pizzo-et-al:15}
{Pizzo}, V.~J., C.~{de Koning}, M.~{Cash}, G.~{Millward}, D.~A. {Biesecker},
  L.~{Puga}, M.~{Codrescu}, and D.~{Odstrcil}.
\newblock {Theoretical basis for operational ensemble forecasting of coronal
  mass ejections}.
\newblock \emph{Space Weather}, \textbf{13}, 676--697, 2015.
\newblock 10.1002/2015SW001221.

\bibitem[{{Prein} et~al.(2017){Prein}, {Rasmussen}, and
  {Stephens}}]{prein-et-al:17}
{Prein}, A.~F., R.~{Rasmussen}, and G.~{Stephens}.
\newblock {Challenges and Advances in Convection-Permitting Climate Modeling}.
\newblock \emph{Bulletin of the American Meteorological Society}, \textbf{98},
  1027--1030, 2017.
\newblock 10.1175/BAMS-D-16-0263.1.

\bibitem[{{Pulkkinen} et~al.(2013){Pulkkinen}, {Rast{\"a}Tter}, {Kuznetsova},
  {Singer}, {Balch} et~al.}]{pulkkinen-et-al:13}
{Pulkkinen}, A., L.~{Rast{\"a}Tter}, M.~{Kuznetsova}, H.~{Singer}, C.~{Balch},
  et~al.
\newblock {Community-wide validation of geospace model ground magnetic field
  perturbation predictions to support model transition to operations}.
\newblock \emph{Space Weather}, \textbf{11}, 369--385, 2013.
\newblock 10.1002/swe.20056.

\bibitem[{{Rast{\"a}tter} et~al.(2014){Rast{\"a}tter}, {T{\'o}th},
  {Kuznetsova}, and {Pulkkinen}}]{rastaetter-et-al:14}
{Rast{\"a}tter}, L., G.~{T{\'o}th}, M.~M. {Kuznetsova}, and A.~A. {Pulkkinen}.
\newblock {CalcDeltaB: An efficient postprocessing tool to calculate
  ground-level magnetic perturbations from global magnetosphere simulations}.
\newblock \emph{Space Weather}, \textbf{12}, 553--565, 2014.
\newblock 10.1002/2014SW001083.

\bibitem[{{Reinert} et~al.(2017){Reinert}, {Prill}, {Frank}, and
  {Z\"{a}ngl}}]{reinert-et-al:17}
{Reinert}, D., F.~{Prill}, H.~{Frank}, and G.~{Z\"{a}ngl}.
\newblock ICON Database Reference Manual.
\newblock \emph{Tech. rep.}, Deutscher Wetterdienst, Offenbach, 2017.
\newblock 10.5676/DWD\_pub/nwv/icon\_1.1.12.

\bibitem[{{Reiss} et~al.(2016){Reiss}, {Temmer}, {Veronig}, {Nikolic},
  {Vennerstrom}, {Sch{\"o}ngassner}, and {Hofmeister}}]{reiss-et-al:16}
{Reiss}, M.~A., M.~{Temmer}, A.~M. {Veronig}, L.~{Nikolic}, S.~{Vennerstrom},
  F.~{Sch{\"o}ngassner}, and S.~J. {Hofmeister}.
\newblock {Verification of high-speed solar wind stream forecasts using
  operational solar wind models}.
\newblock \emph{Space Weather}, \textbf{14}, 495--510, 2016.
\newblock 10.1002/2016SW001390.

\bibitem[{{Riley} et~al.(2018){Riley}, {Baker}, {Liu}, {Verronen}, {Singer},
  and {G{\"u}del}}]{riley-et-al:18}
{Riley}, P., D.~{Baker}, Y.~D. {Liu}, P.~{Verronen}, H.~{Singer}, and
  M.~{G{\"u}del}.
\newblock {Extreme Space Weather Events: From Cradle to Grave}.
\newblock \emph{Space Science Reviews}, \textbf{214}, 21, 2018.
\newblock 10.1007/s11214-017-0456-3.

\bibitem[{{Sch{\"a}r} et~al.(2002){Sch{\"a}r}, {Leuenberger}, {Fuhrer},
  {L{\"u}thi}, and {Girard}}]{schaer-et-al:02}
{Sch{\"a}r}, C., D.~{Leuenberger}, O.~{Fuhrer}, D.~{L{\"u}thi}, and
  C.~{Girard}.
\newblock {A New Terrain-Following Vertical Coordinate Formulation for
  Atmospheric Prediction Models}.
\newblock \emph{Monthly Weather Review}, \textbf{130}, 2459, 2002.
\newblock 10.1175/1520-0493(2002)130<2459:ANTFVC>2.0.CO;2.

\bibitem[{Schmidt et~al.(2017)Schmidt, Bader, Donner, Elsaesser, Golaz, Hannay,
  Molod, Neale, and Saha}]{schmidt-et-al:17}
Schmidt, G.~A., D.~Bader, L.~J. Donner, G.~S. Elsaesser, J.-C. Golaz,
  C.~Hannay, A.~Molod, R.~B. Neale, and S.~Saha.
\newblock Practice and philosophy of climate model tuning across six US
  modeling centers.
\newblock \emph{Geoscientific Model Development}, \textbf{10}(9), 3207--3223,
  2017.
\newblock 10.5194/gmd-10-3207-2017.

\bibitem[{{Schrijver} et~al.(2015){Schrijver}, {Kauristie}, {Aylward},
  {Denardini}, {Gibson} et~al.}]{schrijver-et-al:15}
{Schrijver}, C.~J., K.~{Kauristie}, A.~D. {Aylward}, C.~M. {Denardini}, S.~E.
  {Gibson}, et~al.
\newblock {Understanding space weather to shield society: A global road map for
  2015-2025 commissioned by COSPAR and ILWS}.
\newblock \emph{Advances in Space Research}, \textbf{55}, 2745--2807, 2015.
\newblock 10.1016/j.asr.2015.03.023, \eprint{1503.06135}.

\bibitem[{{Schunk} et~al.(2014){Schunk}, {Scherliess}, {Eccles}, {Gardner},
  {Sojka} et~al.}]{schunk-et-al:14}
{Schunk}, R.~W., L.~{Scherliess}, V.~{Eccles}, L.~C. {Gardner}, J.~J. {Sojka},
  et~al.
\newblock {Ensemble Modeling with Data Assimilation Models: A New Strategy for
  Space Weather Specifications, Forecasts, and Science}.
\newblock \emph{Space Weather}, \textbf{12}, 123--126, 2014.
\newblock 10.1002/2014SW001050.

\bibitem[{{Shaw} and {Daescu}(2017)}]{shaw-daescu:17}
{Shaw}, J.~A., and D.~N. {Daescu}.
\newblock {Sensitivity of the model error parameter specification in
  weak-constraint four-dimensional variational data assimilation}.
\newblock \emph{Journal of Computational Physics}, \textbf{343}, 115--129,
  2017.
\newblock 10.1016/j.jcp.2017.04.050.

\bibitem[{Shimokawabe et~al.(2014)Shimokawabe, Aoki, and
  Onodera}]{shimokawabe-et-al:14}
Shimokawabe, T., T.~Aoki, and N.~Onodera.
\newblock High-productivity Framework on GPU-rich Supercomputers for
  Operational Weather Prediction Code ASUCA.
\newblock In Proceedings of the International Conference for High Performance
  Computing, Networking, Storage and Analysis, SC '14, 251--261. IEEE Press,
  Piscataway, NJ, USA, 2014.
\newblock ISBN 978-1-4799-5500-8.
\newblock 10.1109/SC.2014.26.

\bibitem[{{Sibley} et~al.(2012){Sibley}, {Biesecker}, {Millward}, and
  {Gibbs}}]{sibley-et-al:12}
{Sibley}, A., D.~{Biesecker}, G.~{Millward}, and M.~{Gibbs}.
\newblock {Space weather, impacts and forecasting: an overview}.
\newblock \emph{Weather}, \textbf{67}, 115--120, 2012.
\newblock 10.1002/wea.1915.

\bibitem[{{Simmons} et~al.(2016){Simmons}, {Fellous}, {Ramaswamy}, {Trenberth},
  {Asrar} et~al.}]{simmons-et-al:16}
{Simmons}, A., J.-L. {Fellous}, V.~{Ramaswamy}, K.~{Trenberth}, G.~{Asrar},
  et~al.
\newblock {Observation and integrated Earth-system science: A roadmap for
  2016-2025}.
\newblock \emph{Advances in Space Research}, \textbf{57}, 2037--2103, 2016.
\newblock 10.1016/j.asr.2016.03.008.

\bibitem[{{Singh} et~al.(2010){Singh}, {Siingh}, and {Singh}}]{singh-et-al:10}
{Singh}, A.~K., D.~{Siingh}, and R.~P. {Singh}.
\newblock {Space Weather: Physics, Effects and Predictability}.
\newblock \emph{Surveys in Geophysics}, \textbf{31}, 581--638, 2010.
\newblock 10.1007/s10712-010-9103-1.

\bibitem[{{Siscoe}(2007)}]{sisco:07}
{Siscoe}, G.
\newblock {Space weather forecasting historically viewed through the lens of
  meteorology}.
\newblock In V.~{Bothmer} and I.~A. {Daglis}, eds., Space Weather- Physics and
  Effects, chap.~2, 5--30. Springer, 2007.
\newblock 10.1007/978-3-540-34578-7\_2.

\bibitem[{{Siuta} et~al.(2016){Siuta}, {West}, {Modzelewski}, {Schigas}, and
  {Stull}}]{siuta-et-al:17}
{Siuta}, D., G.~{West}, H.~{Modzelewski}, R.~{Schigas}, and R.~{Stull}.
\newblock {Viability of Cloud Computing for Real-Time Numerical Weather
  Prediction}.
\newblock \emph{Weather and Forecasting}, \textbf{31}, 1985--1996, 2016.
\newblock 10.1175/WAF-D-16-0075.1.

\bibitem[{{Steenburgh} et~al.(2014){Steenburgh}, {Biesecker}, and
  {Millward}}]{steenburgh-et-al:14}
{Steenburgh}, R.~A., D.~A. {Biesecker}, and G.~H. {Millward}.
\newblock {From Predicting Solar Activity to Forecasting Space Weather:
  Practical Examples of Research-to-Operations and Operations-to-Research}.
\newblock \emph{Solar Physics}, \textbf{289}, 675--690, 2014.
\newblock 10.1007/s11207-013-0308-6, \eprint{1305.2791}.

\bibitem[{{Szintai} et~al.(2009){Szintai}, {Kaufmann}, and
  {Rotach}}]{szintai-et-al:09}
{Szintai}, B., P.~{Kaufmann}, and M.~W. {Rotach}.
\newblock {Deriving turbulence characteristics from the COSMO numerical weather
  prediction model for dispersion applications}.
\newblock \emph{Advances in Science and Research}, \textbf{3}, 79--84, 2009.
\newblock 10.5194/asr-3-79-2009.

\bibitem[{{Taylor} et~al.(2012){Taylor}, {Stouffer}, and
  {Meehl}}]{taylor-et-al:12}
{Taylor}, K.~E., R.~J. {Stouffer}, and G.~A. {Meehl}.
\newblock {An Overview of CMIP5 and the Experiment Design}.
\newblock \emph{Bull. Amer. Meteor. Soc.}, \textbf{93}, 485--498, 2012.
\newblock 10.1175/BAMS-D-11-00094.1.

\bibitem[{{T{\'o}th} et~al.(2005){T{\'o}th}, {Sokolov}, {Gombosi}, {Chesney},
  {Clauer} et~al.}]{toth-et-al:05}
{T{\'o}th}, G., I.~V. {Sokolov}, T.~I. {Gombosi}, D.~R. {Chesney}, C.~R.
  {Clauer}, et~al.
\newblock {Space Weather Modeling Framework: A new tool for the space science
  community}.
\newblock \emph{Journal of Geophysical Research (Space Physics)}, \textbf{110,
  A9}, 2005.
\newblock 10.1029/2005JA011126.

\bibitem[{{T{\'o}th} et~al.(2012){T{\'o}th}, {van der Holst}, {Sokolov}, {De
  Zeeuw}, {Gombosi} et~al.}]{toth-et-al:12}
{T{\'o}th}, G., B.~{van der Holst}, I.~V. {Sokolov}, D.~L. {De Zeeuw}, T.~I.
  {Gombosi}, et~al.
\newblock {Adaptive numerical algorithms in space weather modeling}.
\newblock \emph{Journal of Computational Physics}, \textbf{231}, 870--903,
  2012.
\newblock 10.1016/j.jcp.2011.02.006.

\bibitem[{{Tr\'{e}molet}(2006)}]{tremolet:06}
{Tr\'{e}molet}, Y.
\newblock {Accounting for an imperfect model in 4D-Var}.
\newblock \emph{Quarterly Journal of the Royal Meteorological Society},
  \textbf{132}, 2483--2504, 2006.
\newblock 10.1256/qj.05.224.

\bibitem[{{Tsagouri} et~al.(2013){Tsagouri}, {Belehaki}, {Bergeot}, {Cid},
  {Delouille} et~al.}]{tsagouri-et-al:13}
{Tsagouri}, I., A.~{Belehaki}, N.~{Bergeot}, C.~{Cid}, V.~{Delouille}, et~al.
\newblock {Progress in space weather modeling in an operational environment}.
\newblock \emph{Journal of Space Weather and Space Climate}, \textbf{3}(27),
  A17, 2013.
\newblock 10.1051/swsc/2013037.

\bibitem[{Vanderbauwhede and Takemi(2016)}]{vanderbauwhede-takemi:16}
Vanderbauwhede, W., and T.~Takemi.
\newblock An analysis of the feasibility and benefits of GPU/multicore
  acceleration of the Weather Research and Forecasting model.
\newblock \emph{Concurrency and Computation: Practice and Experience},
  \textbf{28}(7), 2052--2072, 2016.
\newblock 10.1002/cpe.3522.

\bibitem[{{Wang} et~al.(2016){Wang}, {Rosen}, {Tsurutani}, {Verkhoglyadova},
  {Meng}, and {Mannucci}}]{wang-et-al-2:16}
{Wang}, C., I.~G. {Rosen}, B.~T. {Tsurutani}, O.~P. {Verkhoglyadova},
  X.~{Meng}, and A.~J. {Mannucci}.
\newblock {Statistical characterization of ionosphere anomalies and their
  relationship to space weather events}.
\newblock \emph{Journal of Space Weather and Space Climate}, \textbf{6}(27),
  A5, 2016.
\newblock 10.1051/swsc/2015046.

\bibitem[{{Watermann} et~al.(2009){Watermann}, {Wintoft}, {Sanahuja}, {Saiz},
  {Poedts} et~al.}]{watermann-et-al2:09}
{Watermann}, J., P.~{Wintoft}, B.~{Sanahuja}, E.~{Saiz}, S.~{Poedts}, et~al.
\newblock {Models of Solar Wind Structures and Their Interaction with the
  Earth's Space Environment}.
\newblock \emph{Space Science Review}, \textbf{147}, 233--270, 2009.
\newblock 10.1007/s11214-009-9494-9.

\bibitem[{{Welling} et~al.(2017){Welling}, {Anderson}, {Crowley}, {Pulkkinen},
  and {Rast{\"a}tter}}]{welling-et-al:17}
{Welling}, D.~T., B.~J. {Anderson}, G.~{Crowley}, A.~A. {Pulkkinen}, and
  L.~{Rast{\"a}tter}.
\newblock {Exploring predictive performance: A reanalysis of the geospace model
  transition challenge}.
\newblock \emph{Space Weather}, \textbf{15}, 192--203, 2017.
\newblock 10.1002/2016SW001505.

\bibitem[{{Zhang} and {Pu}(2010)}]{zhang-pu:10}
{Zhang}, H., and Z.~{Pu}.
\newblock {Beating the Uncertainties: Ensemble Forecasting and Ensemble-Based
  Data Assimilation inModern Numerical Weather Prediction}.
\newblock \emph{Advances in Meteorology}, \textbf{432160}, 10, 2010.
\newblock 10.1155/2010/432160.

\bibitem[{{Zhao} and {Dryer}(2014)}]{zhao-dryer:14}
{Zhao}, X., and M.~{Dryer}.
\newblock {Current status of CME/shock arrival time prediction}.
\newblock \emph{Space Weather}, \textbf{12}, 448--469, 2014.
\newblock 10.1002/2014SW001060.

\end{thebibliography}




\end{document}